\magnification \magstep1
\raggedbottom
\openup 2\jot
\voffset6truemm
\def\II{{\rm 1\!\hskip-1pt I}}
\centerline {\bf ON ELLIPTICITY AND GAUGE INVARIANCE IN}
\centerline {\bf EUCLIDEAN QUANTUM GRAVITY}
\vskip 1cm
\leftline {Ivan G. Avramidi$^{1}$ and Giampiero Esposito$^{2}$}
\vskip 0.3cm
\leftline {\it ${ }^{1}$Department of Mathematics, 
The University of Iowa,}
\leftline {\it 14 MacLean Hall, Iowa City, IA 52242-1419, USA}
\leftline {\it ${ }^{2}$Istituto Nazionale di Fisica Nucleare,
Sezione di Napoli,}
\leftline {\it Mostra d'Oltremare Padiglione 20, 80125 Napoli,
Italy}
\vskip 1cm
\noindent
{\bf Abstract}.
Invariance principles determine many key properties in quantum
field theory, including, in particular, the appropriate form of
the boundary conditions. A crucial consistency check is the proof
that the resulting boundary-value problem is strongly elliptic.
In Euclidean quantum gravity, the appropriate principle seems to
be the invariance of boundary conditions under infinitesimal
diffeomorphisms on metric perturbations, and hence their BRST
invariance. However, if the operator on metric perturbations is
then chosen to be of Laplace type, the boundary-value problem for
the quantized gravitational field fails to be strongly elliptic.
A detailed proof is presented, and the corresponding open problems
are discussed.
\vskip 3cm
\noindent
Paper prepared for the Conference {\it Trends in Mathematical 
Physics}, The University of Tennessee, Knoxville,
October 14-17, 1998.
\vskip 100cm
A very important property that emerges 
from the work in the recent literature$^{1-4}$ is that the
request of complete gauge invariance of the boundary conditions
is sufficient to determine the most general form of local 
boundary-value problem for operators of Laplace type. It is 
therefore quite crucial to understand whether this scheme can be
applied to Euclidean quantum gravity in such a way that the 
ellipticity of the theory is not spoiled. Indeed, despite the
lack of a rigorous mathematical framework for a quantization of
the gravitational field in the Euclidean regime, several good
motivations exist for continuing such an investigation.$^{2}$ In
particular, we would like to stress what follows.
\vskip 0.3cm
\noindent
(i) Quantum gravity via Euclidean path integrals makes it possible
to deal with a framework where, quite naturally, partition functions
are defined. This is, in turn, crucial if one wants to ``combine"
quantum theory, whose predictions are of statistical nature, with
general relativity. 
\vskip 0.3cm
\noindent
(ii) Spectral geometry plays a crucial role if one tries to get a
thorough understanding of the one-loop semiclassical approximation.
In a space-time approach,$^{2}$ such an approximation provides the
``bridge" between the classical world and the as yet unknown full
quantum theory (via path integrals). At some stage, any alternative
approach to quantum gravity should be able to make contact with
what one knows from a perturbative evaluation of transition amplitudes.
\vskip 0.3cm
\noindent
(iii) So much has been learned from the heat-kernel approach to index
theory and to the theory of eigenvalues in Riemannian geometry, that
any new result may have a non-trivial impact on Euclidean quantum
gravity. If one is interested in the most fundamental structures, it
is encouraging to realize that a framework exists where some problems
become well posed. From this point of view, one has to go ahead as
long as possible with Euclidean methods, and the problem is open of
whether one should regard the elliptic boundary-value problems as the
most fundamental tool. 

In our investigation, following Refs. 3,4, we first have to
describe the vector bundle $V$ over the compact Riemannian manifold
$(M,g)$, the ghost operator and the gauge-field operator with their
leading symbols, the projector and the endomorphisms occurring in the
boundary operator. Indeed, in Euclidean quantum gravity, $V$ is the
vector bundle of symmetric rank-two tensor fields over $M$, with 
bundle metric
$$
E^{ab \; cd}=g^{a(c} \; g^{d)b}+\alpha g^{ab}g^{cd},
\eqno (1)
$$
which can be inverted whenever $\alpha \not = -{1\over m}$. The 
generator $\cal R$ of infinitesimal gauge transformations is the Lie
derivative of the symmetric rank-two tensor field $\varphi$ along the
vector field hereafter denoted by $\varepsilon$, i.e.
$$
\Bigr({\cal R}\varepsilon \Bigr)_{ab} \equiv 
(L_{\varepsilon}\varphi)_{ab}=\sqrt{2} \; \nabla_{(a} \; 
\varepsilon_{b)},
\eqno (2)
$$
where $\sqrt{2}$ is a normalization factor.
The corresponding adjoint generator, say $\overline {\cal R}$,
has an action defined by
$$
\Bigr({\overline {\cal R}}\varphi \Bigr)_{a} \equiv 
-\sqrt{2} \; E_{a}^{\; bcd}\nabla_{b}\varphi_{cd}.
\eqno (3)
$$
Thus, the ghost operator $L \equiv {\overline {\cal R}}{\cal R}$
turns out to have the leading symbol$^{3}$ 
$$
\sigma_{L}(L;\xi)\varepsilon_{a}=2E_{a}^{\; bcd}\xi_{b}\xi_{c}
\varepsilon_{d}=\left[\delta_{a}^{b}{|\xi |}^{2}+(1+2\alpha)
\xi_{a}\xi^{b}\right]\varepsilon_{b}.
\eqno (4)
$$
To turn $L$ into an operator of Laplace type, which is part of our
programme, it is therefore necessary to set $\alpha=-{1\over 2}$. 
To obtain a positive-definite leading symbol for $L$, it would 
instead be enough to require that $\alpha > -1$. Moreover, the
leading symbol of the operator $H \equiv {\cal R}{\overline {\cal R}}$
reads$^{3}$ 
$$
\sigma_{L}(H;\xi)\varphi_{ab}=2\xi_{(a} \; E_{b)}^{\; \; cde}
\xi_{c}\varphi_{de}=\left[2 \xi_{(a} \; \delta_{b)}^{(c} \;
\xi^{d)} + 2\alpha \xi_{a}\xi_{b}g^{cd}\right]\varphi_{cd}.
\eqno (5)
$$
The gauge-invariant operator $\bigtriangleup$ resulting from the
Einstein--Hilbert action is well known to have leading symbol$^{3}$
$$ \eqalignno{
\; & \sigma_{L}(\bigtriangleup;\xi)\varphi_{ab}=\left[
\delta_{(a}^{(c} \; \delta_{b)}^{d)} {|\xi|}^{2}
+\xi_{a}\xi_{b}g^{cd} \right . \cr
& \left . -2\xi_{(a} \; \delta_{b)}^{(c} \; \xi^{d)}
+{(1+2\alpha)\over (1+m\alpha)}g_{ab}(\xi^{c}\xi^{d}-g^{cd})
\right]\varphi_{cd}.
&(6)\cr}
$$
By virtue of Eqs. (5) and (6), the gauge-field operator 
$F \equiv \bigtriangleup + H$ is of Laplace type only if
$\alpha=-{1\over 2}$, just as for the ghost. Hereafter, we shall
instead consider for a while an operator of Laplace type, say $P$,
acting on symmetric rank-two tensor fields with fibre metric (1)
for arbitrary $\alpha$.

We are interested in self-adjoint projection operators 
$\Pi_{ab}^{\; \; \; cd}$ with respect to the bundle metric (1). On 
denoting by $N^{a}$ the inward-pointing normal vector field to
the boundary of $M$, and by $q_{a}^{\; b} \equiv 
\delta_{a}^{\; b}-N_{a}N^{b}$ the
tensor field which projects vectors onto $\partial M$, 
one obtains the desired projector in the form$^{3}$ 
$$
\Pi_{ab}^{\; \; \; cd} \equiv q_{(a}^{\; \; c} \; q_{b)}^{\; \; d}
-{\alpha \over (\alpha+1)}N_{a}N_{b}q^{cd}.
\eqno (7)
$$
In the applications, it will be useful to bear in mind that the
fibre trace of $\Pi_{ab}^{\; \; \; cd}$ is equal to
${m\over 2}(m-1)$. The local boundary conditions for our operator
of Laplace type can be originally written in the form 
$$
\Bigr[\Pi_{ab}^{\; \; \; cd} \; \varphi_{cd}\Bigr]_{\partial M}=0,
\eqno (8)
$$
$$
\Bigr[E^{ab \; cd}\nabla_{b}\varphi_{cd}\Bigr]_{\partial M}=0.
\eqno (9)
$$
Equation (9) involves both normal and tangential derivatives of
$\varphi$, so that (8) and (9) are eventually re-expressed as
$$
\pmatrix{\Pi & 0 \cr \Lambda & \II-\Pi \cr}
\pmatrix{[\varphi]_{\partial M} \cr 
[\nabla_{N}\varphi]_{\partial M} \cr}=0,
\eqno (10)
$$
with
$$
\Lambda \equiv (\II-\Pi)\left[{1\over 2}\Bigr(\Gamma^{i}
{\widehat \nabla}_{i}+{\widehat \nabla}_{i}\Gamma^{i}\Bigr)
+S \right](\II-\Pi).
\eqno (11)
$$
Here, in particular,$^{3}$ 
$$
\Gamma_{ab}^{i \; \; cd} \equiv -{1\over (1+\alpha)}N_{a}N_{b}
e^{i(c} \; N^{d)}+N_{(a} \; e_{b)}^{\; \; i} N^{c}N^{d},
\eqno (12)
$$
where $e^{i}_{a}$ is a local tangent frame on $\partial M$,
and $S$ is an endomorphism.
Thus, on denoting by $\zeta_{i}$ the cotangent vectors on the
boundary: $\zeta_{j} \in T^{*}({\partial M})$, the matrix 
$T \equiv \Gamma^{j}\zeta_{j}$ occurring in the condition for
strong ellipticity$^{3}$ of the boundary-value problem $(P,B_{P})$
reads, in our case,
$$
T=-{1\over (1+\alpha)}p_{1}+p_{2},
\eqno (13)
$$
where, having defined $\zeta_{a} \equiv e_{a}^{\; i}\zeta_{i}$,
one has$^{3}$ 
$$
(p_{1})_{ab}^{\; \; \; cd} \equiv N_{a}N_{b}
\zeta^{(c} \; N^{d)},
\eqno (14)
$$
$$
(p_{2})_{ab}^{\; \; \; cd} \equiv N_{(a} \; \zeta_{b)}
N^{c}N^{d}.
\eqno (15)
$$
The matrix $T$ is annihilated by the projector $\Pi$ on the left 
and on the right, as is necessary to be consistent with the 
desired scheme. Moreover, on studying the products of the matrices
$p_{1}$ and $p_{2}$, one is led to consider further projectors 
given by$^{3,4}$ 
$$
\rho_{ab}^{\; \; \; cd} \equiv {2\over |\zeta|^{2}}
N_{(a} \; \zeta_{b)} N^{(c} \; \zeta^{d)},
\eqno (16)
$$
$$
p_{ab}^{\; \; \; cd} \equiv N_{a}N_{b}N^{c}N^{d}.
\eqno (17)
$$
One then finds
$$
p_{1}p_{2}={1\over 2}{|\zeta|}^{2}p,
\eqno (18)
$$
$$
p_{2}p_{1}={1\over 2}{|\zeta|}^{2}\rho,
\eqno (19)
$$
where the matrices $p_{1}$ and $p_{2}$ are nilpotent:
$(p_{1})^{2}=(p_{2})^{2}=0$, whilst $\rho$ and $p$ are mutually
orthogonal: $p \rho=\rho p=0$. Thus, on defining
$$
\tau \equiv {1\over \sqrt{2(1+\alpha)}}|\zeta|,
\eqno (20)
$$
the square of the matrix $T$ takes the simple and elegant form
$$
T^{2}=-\tau^{2}(p+\rho).
\eqno (21)
$$
General formulae for even and odd powers of $T$ are also useful, i.e.
$$
T^{2n}=(i\tau)^{2n}(p+\rho),
\eqno (22)
$$
$$
T^{2n+1}=(i\tau)^{2n}T.
\eqno (23)
$$
The nilpotent matrices $p_{1}$ and $p_{2}$ have vanishing trace,
whereas the projectors $p$ and $\rho$ have unit trace. Hence
one finds 
$$
{\rm tr} \; T^{2n}=2(i\tau)^{2n},
\eqno (24)
$$
$$
{\rm tr} \; T^{2n+1}={\rm tr} \; T=0,
\eqno (25)
$$
which implies, in turn, that for any function, say $f$, analytic in
the region $|z| \leq \tau$, one has$^{3,4}$
$$ \eqalignno{
\; & f(T)=f(0)[\II-p-\rho]
+{1\over 2}\Bigr[f(i\tau)+f(-i \tau)\Bigr](p+\rho) \cr
&+{1\over 2i \zeta}\Bigr[f(i\tau)-f(-i\tau)\Bigr]T,
&(26)\cr}
$$
$$
{\rm tr} \; f(T)=\left[{m(m+1)\over 2}-2 \right]f(0)
+f(i\tau)+f(-i \tau).
\eqno (27)
$$
Thus, the matrix $T$ has eigenvalues $0, i\tau$ and $-i \tau$, and
the matrix $T^{2}$, for non-vanishing $\zeta_{j}$, has eigenvalues
$0$ and $-\tau^{2}=-{|\zeta|^{2}\over 2(1+\alpha)}$. In other words,
for $|\zeta| \not = 0$, the matrix $\Bigr(T^{2}+|\zeta|^{2} \II
\Bigr)$ is positive-definite, so that the boundary-value problem
$(P,B_{P})$ is strongly elliptic with respect to 
${\bf C}-{\bf R}_{+}$, if and only if$^{3}$ 
$$
-{1\over 2(1+\alpha)}+1 > 0,
\eqno (28)
$$
which implies $\alpha > -{1\over 2}$. But we know from Eqs. (5)
and (6) that, in one-loop Euclidean quantum gravity, the gauge-field
operator can be of Laplace type if and only if $\alpha=-{1\over 2}$. 
Thus, we have proved, as a corollary, that the gauge-invariant 
boundary conditions (8) and (9) make it impossible to achieve 
strong ellipticity with respect to ${\bf C}-{\bf R}_{+}$ in Euclidean
quantum gravity.$^{3,4}$ 

To appreciate which non-trivial properties result from the lack of
strong ellipticity we are now going to prove that the heat-kernel
diagonal, although well defined, has a non-integrable behaviour
when the geodesic distance from the boundary, say $r$, tends to
zero (unlike the smooth behaviour which is obtained when strong
ellipticity is not violated). Indeed, from the work in Ref. 3, one knows
that the fibre trace of the zeroth-order heat-kernel diagonal reads
$$
{\rm tr}_{V}U_{0}(x,x;t)=(4\pi t)^{-m/2}\left[C_{0}
+C_{1}e^{-r^{2}/t}+J(r/\sqrt{t})\right],
\eqno (29)
$$
where
$$
C_{0} \equiv {\rm dim}V={m(m+1)\over 2},
\eqno (30)
$$
$$
C_{1} \equiv {\rm tr}_{V}(\II-2\Pi)=-{m(m-3)\over 2},
\eqno (31)
$$
$$ \eqalignno{
\; & J(z) \equiv -2 \int_{{\bf R}^{m-1}}
{d\zeta_{1}... d\zeta_{m-1}\over \pi^{(m-1)/2}} \cr
& \int_{\gamma}{d\omega \over \sqrt{\pi}}
e^{-|\zeta|^{2}-\omega^{2}+2i \omega z}{\rm tr}_{V}
\; \Gamma^{j}\zeta_{j}(\omega \II + \Gamma^{k}\zeta_{k})^{-1}.
&(32)\cr}
$$
By virtue of (27), with $\tau=|\zeta|$ for $\alpha=-{1\over 2}$, 
one finds
$$
J(z)=-4 \int_{{\bf R}^{m-1}}{d\zeta \over \pi^{(m-1)/2}}
\int_{\gamma}{d\omega \over \sqrt{\pi}}
e^{-|\zeta|^{2}-\omega^{2}+2i \omega z}
{|\zeta|^{2}\over \omega^{2}+|\zeta|^{2}},
\eqno (33)
$$
where the contour $\gamma$ comes from $-\infty+i \varepsilon$, goes
around the pole $\omega=i |\zeta|$ in the clockwise direction and then
goes to $+\infty+i \varepsilon$. The integral with respect to $\omega$
is evaluated with the help of the well known formula
$$
\int_{\gamma}f(\omega)d\omega=
-2\pi i \Bigr[{\rm Res} \; f(\omega)\Bigr]_{\omega=i |\zeta|}
+\int_{-\infty}^{\infty}f(\omega)d\omega ,
\eqno (34)
$$
and the integrals with respect to $\zeta_{1},...,\zeta_{m-1}$
(see (32)) can be reduced to Gaussian integrals. This method leads
to the result$^{3}$
$$
J(z)=2(m-1)z^{-m}\Gamma(m/2,z^{2}),
\eqno (35)
$$
where we have used the incomplete $\Gamma$-function
$\Gamma(a,x) \equiv \int_{x}^{\infty}u^{a-1}e^{-u}du$. The function
$J$ is singular as $z \rightarrow 0$, in that
$$
J(z) \sim 2(m-1) \Gamma(m/2)z^{-m},
\eqno (36)
$$
whereas, as $z \rightarrow \infty$, it is exponentially damped, i.e.
$$
J(z) \sim 2(m-1)z^{-2}e^{-z^{2}},
\eqno (37)
$$
as can be seen by using the identity $\Gamma(a,x)=x^{a-1}e^{-x}
+(a-1)\Gamma(a-1,x)$. The singularity at $z=0$ results from the
pole at $\omega=i |\zeta|$ and provides a peculiar difference with
respect to the strongly elliptic case, where all poles lie on the
positive imaginary line with ${\rm Im}(\omega)< i |\zeta|$. The
fibre trace (29) is now re-expressed as$^{3}$
$$
{\rm tr}_{V} \; U_{0}(x,x;t)=(4\pi t)^{-m/2}
\left[C_{0}+C_{1}e^{-r^{2}/t}+2(m-1)(r/\sqrt{t})^{-m}
\Gamma(m/2,r^{2}/t)\right].
\eqno (38)
$$
One can see that the first term in (38) yields the familiar
interior contribution, upon integration over a compact manifold.
Moreover, the boundary terms in (38) are exponentially small
as $t \rightarrow 0^{+}$, if $r$ is fixed. By contrast, on fixing
$t$, one finds, as $r \rightarrow 0$,
$$
{\rm tr}_{V} \; U_{0}(x,x;t) \sim (4\pi)^{-m/2}2(m-1)
\Gamma(m/2)r^{-m}.
\eqno (39)
$$
Remarkably, such a limiting behaviour is independent of $t$ and
is not integrable with respect to $r$ as one approaches the 
boundary, i.e. as $r \rightarrow 0$.$^{3}$ This is an enlightening
example of the undesirable properties resulting from the lack of
strong ellipticity.

Our work, relying on Refs. 3,4, has proved that, upon studying Euclidean
quantum gravity at one-loop level on compact Riemannian manifolds
with smooth boundary, if one tries to have a gauge-field operator
and ghost operator both of Laplace type, with local boundary
operator (see (10) and (11)) and completely gauge-invariant boundary
conditions, the ellipticity of the theory is lost in that the
boundary-value problem for metric perturbations fails to be 
strongly elliptic with respect to ${\bf C}-{\bf R}_{+}$. Since the
properties of ellipticity on the one hand, and invariance under 
infinitesimal diffeomorphisms (on metric perturbations) on the other
hand, are both at the heart of the Euclidean approach to quantum
gravity, the proof of their incompatibility appears to be a matter
of serious concern. At least three alternatives deserve further
investigation. They are as follows.
\vskip 0.3cm
\noindent
(i) The use of smearing functions (e.g. $f:r \rightarrow
f(r)=r^{m}$) in defining the functional trace of $e^{-tF}$, with
$F$ the gauge-field operator $\bigtriangleup+H$.
\vskip 0.3cm
\noindent
(ii) The attempt to prove strong ellipticity when the gauge-field
operator is non-minimal.
\vskip 0.3cm
\noindent
(iii) The attempt to recover strong ellipticity upon studying 
non-local boundary conditions, with the help of the current
progress in the functional calculus of pseudo-differential
boundary problems.$^{5,6}$ 
\vskip 0.3cm
The latter line of investigation, although quite difficult at the
technical level, might lead to a deeper vision,$^{6}$ and finds its
motivation not only in the recent mathematical progress,$^{5}$ but
also in well known properties of quantum field theory, e.g.
the non-local nature of projectors on the gauge-invariant
sub-space of the configuration space.$^{3}$ There is therefore
some intriguing evidence that the most fundamental problems in
spectral geometry and quantum gravity are still waiting for a
proper solution.
\vskip 0.3cm
\leftline {\bf References}
\vskip 0.3cm
\item {[1]}
A. O. Barvinsky, {\it Phys. Lett.}, {\bf B195} (1987) 344. 
\item {[2]}
G. Esposito, A. Yu. Kamenshchik and G. Pollifrone,
{\it Euclidean Quantum Gravity on Manifolds with Boundary},
Fundamental Theories of Physics, Volume 85
(Kluwer, Dordrecht, 1997).
\item {[3]}
I. G. Avramidi and G. Esposito, {\it Gauge Theories on 
Manifolds with Boundary} (HEP-TH 9710048, to appear in
{\it Commun. Math. Phys.}).
\item {[4]}
I. G. Avramidi and G. Esposito, {\it Class. Quantum Grav.},
{\bf 15} (1998) 1141.
\item {[5]}
G. Grubb, {\it Functional Calculus of Pseudo-Differential
Boundary Problems}, Progress in Mathematics, Volume 65 
(Birkh\"{a}user, Boston, 1996).
\item {[6]}
G. Esposito, {\it Non-Local Boundary Conditions in Euclidean
Quantum Gravity} (GR-QC 9806057).

\bye